\documentclass[preprint,showpacs,preprintnumbers,amsmath,amssymb,prb]{revtex4}

\usepackage{graphicx}

\begin{document}

\preprint{PRB}

\title{Superconducting critical current of a single $\mathbf{Cu_2O_4}$-plane in $\mathbf{Bi_2Sr_2CaCu_2O_{8+x}}$ single crystal}

\author{L. X. You}
\email{lixing@mc2.chalmers.se}
\author{A. Yurgens}
\author{D. Winkler}

\affiliation{Quantum Device Physics Laboratory, Department of
Microtechnology and Nanoscience, Chalmers University of
Technology, SE-412 96 G\"oteborg, Sweden}

\date{\today}

\begin{abstract}
By feeding current into the topmost $\mathrm{Cu_2O_4}$-layer
of a mesa etched into the surface of
a $\mathrm{Bi_2Sr_2CaCu_2O_{8+x}}$ (BSCCO) single crystal,
we measured its superconducting critical value from a
sharp upturn or break in current-voltage characteristics of the
mesas. From this,
we estimate the sheet critical current density of a single
$\mathrm{Cu_2O_4}$ plane to be $\sim$\ 0.3-0.7~A/cm at 4.5~K,
corresponding to the bulk current density of $~2-5\ \mathrm{MA/cm^2}$.  These
values are among the largest ever measured for BSCCO single crystals,
thin films and tapes.

\end{abstract}


\maketitle

\section{Introduction}

The naturally layered structure along the $c$-axis of high
temperature superconductors (HTS) is an important feature that
accounts for the large anisotropy of transport and superconducting
properties along and perpendicular to the layers.  It is well
established that the $c$-axis transport results from sequential {\it
tunneling} of charge carriers between the Cu-O planes that turns
into Josephson tunneling below the superconducting critical
temperature $T_c$. This intrinsic Josephson effect was first
discovered about ten years ago\cite{Kleiner:PRL92,Kleiner:PRB94} and
has been a subject of many studies since then \cite{Yurgens:SUST00}.
It is worth noting that no other examples of Josephson tunneling
that occurred in entirely single-crystalline media had been known
until the discovery.

\par

To justify commercial use of HTS materials in power applications,
the critical current densities of prototype cables should be
sufficiently high. The bottleneck of HTS-cable performance is
assumed to be in poor inter-grain connectivity and bad $c$-axis
conductivity of the material while the in-plane ($ab$-)
superconducting critical current densities, $J_{c \parallel ab}$,
are acceptably high. From both a fundamental and an applied point of
view, it is important to know whether the measured $J_{c
\parallel ab}$ reflects the genuine properties of the material. In most cases
however, the measurements of $J_{c \parallel ab}$ have been made on
relatively large single crystals and thin films. Those might include
stacking faults and grain boundaries which obviously limit the
observed $J_{c \parallel ab}$. Moreover, to correctly measure $J_{c
\parallel ab}$  for materials with the highest anisotropy, like
BSCCO, the current should be injected uniformly in every Cu-O plane
across the thickness of the single crystal, while any imbalance will
lead to redistribution of current between the planes thus involving
the out-of-plane ($c$-axis) properties.

\par

In this paper, we present a method of measuring $J_{c
\parallel ab}$  of a \textit{single} $\mathrm{Cu_2O_4}$ (Cu-O) plane in
$\mathrm{Bi_2Sr_2CaCu_2O_{8+x}}$ (BSCCO) utilizing its layered
nature and the presence of the intrinsic Josephson
effect\cite{Kleiner:PRL92,Kleiner:PRB94}. We use an extended
intrinsic Josephson junction (IJJ)\cite{long_junction} and a
non-uniform current bias\cite{Barone} to force the current to flow
along the topmost electrode of the junction, i.e. along the single
Cu-O plane. A distinct feature of the current-voltage ($I-V$)
characteristic of the IJJ marks the moment when this current exceeds
its superconducting critical value within the Cu-O plane.

\par

The thickness of one intrinsic Josephson junction in BSCCO is only
1.5 nm, which explains why it is difficult to isolate and study such
a single junction. As a result, most results on IJJs were obtained
from stacks (mesas) containing many IJJs. There were a few
successful attempts of making stacks enclosing a single IJJ, either
by accident\cite{Yurgens:PRB96} or by using a tricky etching
technique involving {\it in situ} monitoring of the resulting
current-voltage ($I-V$) characteristics of the stack
\cite{Yurgens:APL97}.

\par

By precisely controlling the fabrication parameters, we successfully
made stacks with a low number of IJJs, as well as a single intrinsic
Josephson junction (SIJJ) \cite{You:SUST03,You:JJAP04}. In our
four-probe measurements of SIJJ, a steep upturn of the quasiparticle
branch occurs at a relatively low bias current when the in-plane
current reaches the critical value for a single Cu-O plane. From
this we can estimate the sheet critical current density of a single
Cu-O plane to be $\sim 0.3-0.7$~A/cm at 4.5~K. A back-bending of the
quasiparticle branch followed by a reentrance to the zero-voltage
state is also observed and is explained by significant Joule heating
at higher bias currents.

In stacks with many junctions, the transition of the topmost layer
is seen as a break in the $I-V$ characteristics, most clearly in the
last quasiparticle branch. Observed by several groups, these breaks
have not been elucidated until now.

\section{Sample preparation}

The samples were made from nearly optimally doped BSCCO single
crystals.  The stacks of IJJs were formed by photolithography and
Ar-ion etching in two steps.

\par

First, the single crystal is cleaved and a thin film of gold (20-30
nm) is deposited in order to protect the surface from deterioration
during photolithography patterning. Then, by controlling the process
parameters\cite{You:SUST03,You:JJAP04}, a stack of a certain height
$h \sim 10$ nm is etched into a single crystal.  At last, during the
second Ar-ion etching, a gap between the contacts is formed in the
middle of the stack. The gap is etched to a specified depth $d = h -
n \times 1.5$ nm. The height $h$ and the depth $d$ could also be
verified to the accuracy of one junction ($\sim 1.5$ nm) by
measuring the $I-V$ characteristics of the resulting sample at low
temperature \cite{NJ}.

\begin{figure}
\includegraphics{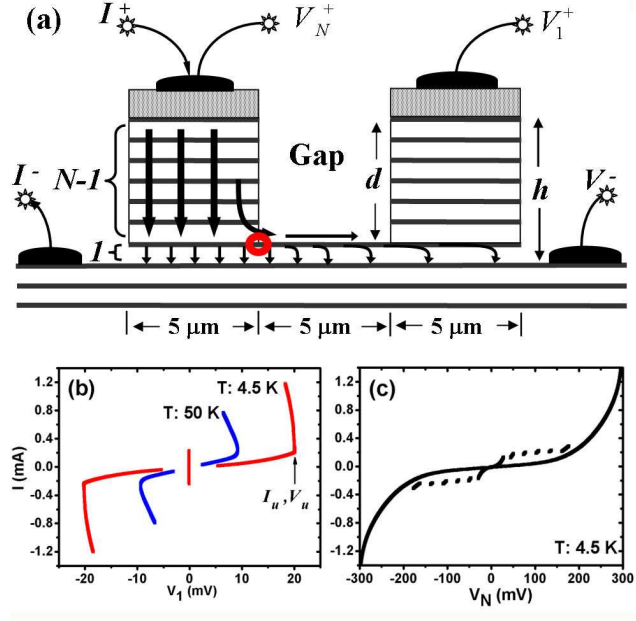}
\caption{(Color online) (a) schematic view of the sample with a SIJJ; (b)
four-probe $I-V$ curves of a single intrinsic Josephson junction
with the voltage measured between $V_1^+$  and $V^-$-contacts at two
temperatures, $T=$ 4.5 and 50 K; (c) three-probe $I-V$ curves of the
sample with the voltage measured between $V_N^+$  and $V^-$ at 4.5
K. Note the voltage jumps corresponding to 7 junctions (the
SIJJ plus all the junctions under the current contact).}
\label{SIJJ}
\end{figure}

In the end, the overall stack acquires a U-shaped form, see
Fig.~\ref{SIJJ}a, with two smaller stacks under the current and
potential contacts sitting on top of the common pedestal which is
only 1.5 nm high. Following this technique, we can make a specified
number of effective junctions enclosed in the base stack, including
the most interesting case of SIJJ in this paper. An account of the
detailed fabrication process is published elsewhere
\cite{You:SUST03,You:JJAP04}.

\par

In the four-terminal measurements (Fig.~\ref{SIJJ}a), two contacts
are on the top of the SIJJ while other two are somewhere else on
the crystal surface outside the original stack\cite{contacts}. With
voltage measured between $V_1^+$ and $V^-$, only the effective
single junction in the base is registered (Fig.~\ref{SIJJ}b).
However, the extra 5-6 junctions underneath the contact stacks may
have a profound effect on the resulting $I-V$ curve due to Joule
heating, as will be discussed below.

\section{surface topography}

In the suggested method, it is important that the top electrode of
the SIJJ is spatially uniform. One may argue however that the Ar-ion
etching is not sufficiently homogeneous to produce an atomically
smooth surfaces after the etching.

\par

\begin{figure}
\includegraphics{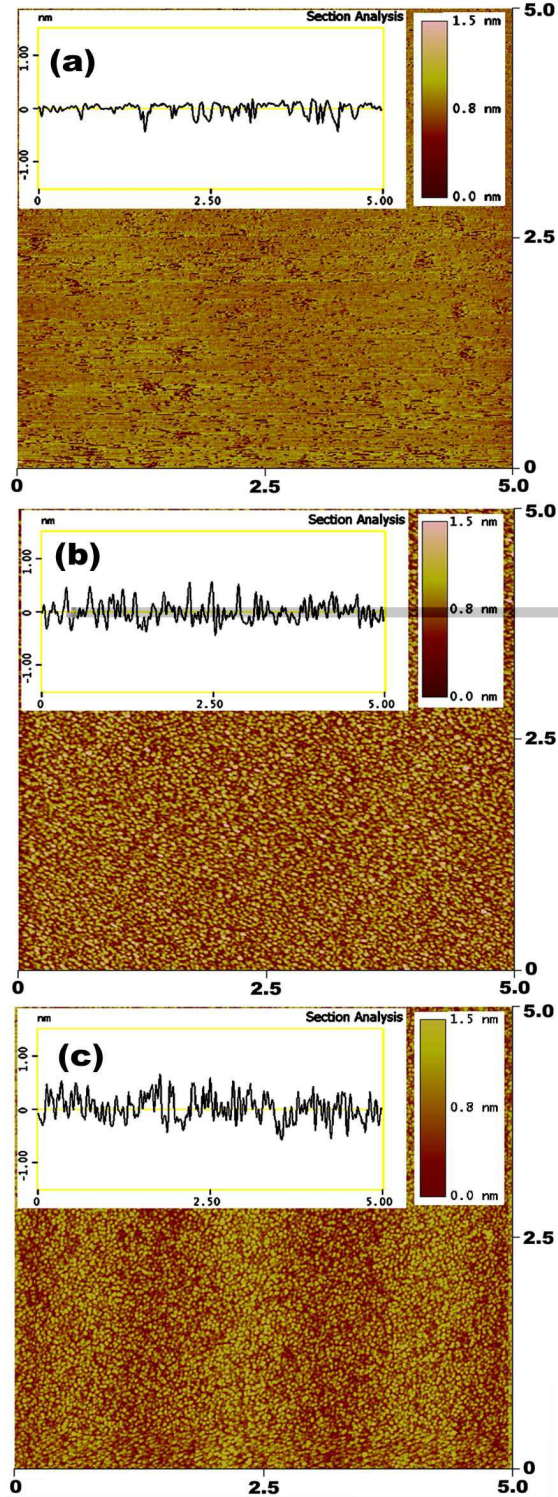}
\caption{AFM analysis of BSCCO surface. (a) a freshly cleaved BSCCO
surface with the rms surface roughness of 0.20 nm; (b) the surface
outside the mesa with the rms surface roughness of 0.26 nm; (c)
BSCCO surface in the middle of the mesa (``gap" region) with the rms
surface roughness of 0.38 nm. The imaged areas are $5 \times 5\ \mu
m^2$}. \label{AFM}
\end{figure}

To examine the surface roughness before and after the etching, the
Atomic Force Microscope (AFM) is used.  Fig.~\ref{AFM} shows the
BSCCO surface of another sample in the gap region and outside it
after etching, as well as the virgin BSCCO surface before the
processing.

It is seen that the rms surface roughness of the etched areas is
0.38 nm at most, to be compared to the roughness of the freshly
cleaved surface of 0.20 nm.

\par

All these values are less than one third of the thickness of the
insulating barrier layers between the Cu-O layers (1.2 nm), which
means that the etched surface of BSCCO is flat enough for forming a
sufficiently uniform top electrode of SIJJ.

\par

The high degree of smoothness of the etched area is partly due to
the low ion energy (230~eV) and beam intensity ($7\times
10^{14}$~s$^{-1}$cm$^{-2}$$\sim 0.11$~mA/cm$^2$) used throughout the
etching process\cite{kamimura:OL02}.

\section{Low-current measurements}

All the measurements were carried out in a liquid-helium dewar. The
temperature was changed by placing the specimen in cold He-vapor
above the liquid-He level.

\par

Fig.~\ref{SIJJ}b shows a typical $I-V$ curve of a SIJJ at two
temperatures where only one quasiparticle branch is seen. A
sharp upturn of the quasiparticle branch is observed at about
$V_u=20$ mV at 4.5 K and bias current $I_u= 0.22$ mA denoted by the
arrow in Fig.~\ref{SIJJ}b. A three-terminal measurement is also shown
in Fig.~\ref{SIJJ}c for comparison, where all IJJs situated in the
small stack under the current contact and the SIJJ are seen.

\par

By examining the current flow through the structure, we argue that
the sharp upturn at ($V_u, I_u$)-point is due to the in-plane
current flowing along the topmost Cu-O plane of the SIJJ exceeding
its critical value. Indeed, when the bias current is spatially
non-uniform for the SIJJ, i.e. is applied to one end, it first tends
to distribute itself over the whole surface of the SIJJ's top
electrode. This means that a finite in-plane superconducting current
must flow between the current and potential electrodes, as depicted
in Fig.~\ref{SIJJ}a. If it exceeds the critical value, the top
electrode becomes resistive thus breaking the connection between the
two parts of the junction. Then, the current will be redistributed
between these parts, dependent on their particular $c$-axis
resistances and the resistance of the ``bridge" between them. The
corresponding voltage measured at the potential contact will be a
function of all the non-linear resistances involved.

\par

The highest in-plane current density is expected to be in the place
where the small mesa under the bias electrode ends and where the
bias current starts to spread itself along the top electrode of the
SIJJ  (marked by a circle in Fig.~\ref{SIJJ}a). Each electrode (the
small mesa $\sim 5 \times 5\ \mathrm{\mu m^2}$) occupies about one
third of the total area of the SIJJ ($\sim 15 \times 5\ \mathrm{\mu
m^2}$). This means that initially, when the top electrode of the
SIJJ is still superconducting, about one third of the bias current
passes along the $c$ axis directly through the area under the
current-bias contact. The remaining two thirds will flow
non-uniformly through the rest of the junction area.

\par

It is easy to understand now that the maximum value of the
$ab$-plane superconducting current is roughly equal to two thirds of
the total bias current. The (sheet) critical current density of a
single Cu-O plane can be estimated to be about 0.3 A/cm after
dividing $2/3I_u$ by the width of the SIJJ (5 $\mathrm{\mu m}$). To
be able to compare this value to the published bulk critical current
densities, we need to take into account the total thickness of SIJJ
(1.5 nm). Hence, the (bulk) in-plane critical current density of
BSCCO  can be  estimated to be about 2.0 $\mathrm{MA/cm^2}$.
Although being quite reasonable, this value is among the largest
ever observed in other $J_{c \parallel ab}$ measurements
\cite{Li:PhysicaC96,Matsushita:SUST98,Latyshev:PhysicaC93,Mizutani:Physica03,
Flahaut:IEEEASC03,Marken:IEEEASC01,Yonemitsu:PhysicaC02,Moriya:SUST02}.

\par

Further proof of the method was obtained from another sample ($18
\times 3\ \mu m^2$) with two top electrodes of different sizes
($S_L$: $9 \times 3\ \mu m^2$ and $S_R$: $5 \times 3\ \mu m^2$). By
passing the current through two different top electrodes, two
different values of $I_u$ ($I_{u:L}=0.45$ mA and $I_{u:R}=0.31$ mA)
for the sharp upturn structure were observed in similar $I-V$
curves. However, considering the sizes of the electrodes, the same
critical current $I_{c \parallel ab}= 0.22$ mA of a single Cu-O
plane was obtained. The corresponding sheet critical current density
of 0.7~A/cm is higher than the value for the former sample, which
can be explained by the spread of $J_c$ from crystal to crystal.

\par

\begin{figure}
\includegraphics{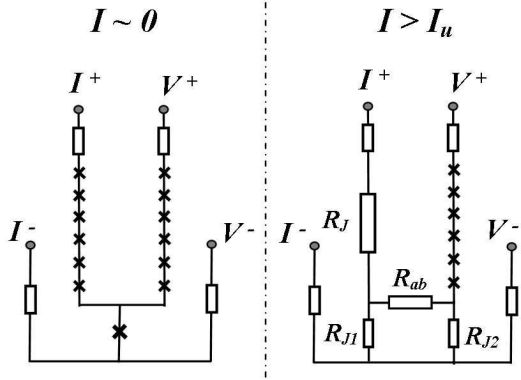} \caption{
A simple equivalent circuit of the junction stack in four-probe
measurement.  The crosses schematically show the intrinsic Josephson
junctions.  Small unmarked resistors represent resistance of
connecting wires and the contact resistances.  In the
quasiparticle-tunneling state, the junctions are represented by the
corresponding non-hysteretic and non-linear resistances.}
\label{circuit}
\end{figure}

A simple equivalent circuit shown in Fig.~\ref{circuit} can be used
to highlight the situation qualitatively.  We model the junction
resistances in the $c$-axis direction by non-linear $V_J(I)$
functions. $V_J(I)$ per junction are deduced from the measured
$V(I)$-dependence corresponding to the last quasiparticle branch in
Fig.~\ref{SIJJ}c by dividing voltage by the number of junctions in
the whole mesa.

In order to model the dynamics of current flow through the SIJJ, we
assume quite arbitrarily the following form of the voltage drop
across the $ab$ plane at the bridge between the two regions:
\begin{equation} \label{eq1}
V_{ab}(I)={R_{ab}I}\left[1+\exp \left(\frac{I_c-I}{\delta I}\right)\right]^{-1}
\end{equation}

$V_{ab}(I)$ of Eq.\ref{eq1} simulates a transition from the
superconducting to the normal state of the top electrode of SIJJ. We
see that $V_{ab}(I)$ is exponentially small when $I < I_c$
(``superconducting state") and is finite and ohmic,
$V_{ab}(I)=R_{ab}I$, when $I > I_c$ (normal state with the
resistance $R_{ab}$). $\delta I$ represents the ``rounding" of the
transition ($\delta I \ll I_c$).

\par

\begin{figure}
\includegraphics{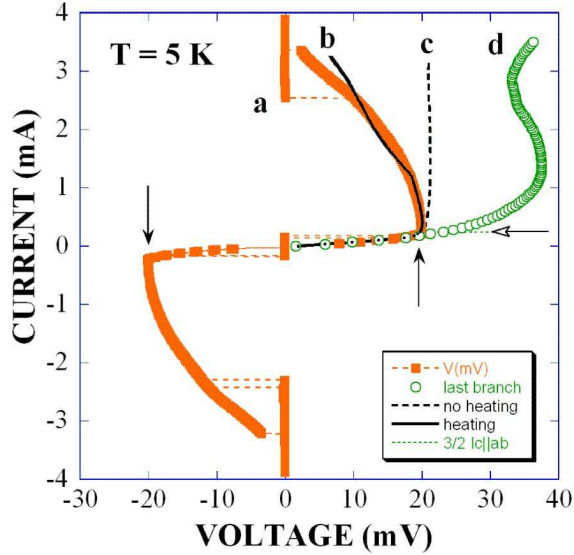} \caption{(Color online) a) The four-probe
current-voltage curve of the single intrinsic Josepson junction at
high bias; b,c) simulated $I-V$ cuves using the simple equivalent
circuit depicted in Fig.~\ref{circuit}.  The horizontal dashed line
and hollow arrow show the value of bias current corresponding to
1.5$I_{c \parallel ab}$. The vertical arrows mark the sharp upturn
of the quasiparticle branch; d) $I-V$ curve assumed for a single
junction when the bias current is uniform. This curve was obtained
from the last quasiparticle branch (see Fig.~\ref{SIJJ}c) by dividing
the voltage by the number of junctions involved.} \label{IVC}
\end{figure}

The equivalent circuit shown in Fig.~\ref{circuit} can be further
used to qualitatively simulate even the resulting current-voltage
curve at small bias.  It is shown in Fig.~\ref{IVC} for the
following parameters: $\delta I = 0.003$ mA, $I_c = 0.15$ mA, and
$R_{ab} = 2$ k$\Omega$.  The latter corresponds to the sheet
resistance of a single Cu-O plane taking the $ab$-resistivity to be
300 $\mathrm{\mu \Omega /cm}$. \cite{Watanabe:PRL97}

The dashed line c) in Fig.~\ref{IVC} mimics the steep upturn of the
measured $I-V$ curve and explains the upturn structure at low bias
current qualitatively well.

\par

It should be also noted that at the ($V_u, I_u$)-point where the
break and the upturn of the quasiparticle branch are seen, the Joule
dissipation and, hence, overheating is quite small and cannot
explain these features. Indeed, from Fig.~\ref{SIJJ}c it follows that
the total dissipation of the whole stack is less than 50 $\mu$W at
$I = I_u$. Using typical values for the thermal resistance
\cite{Yurgens:PRL04}, 40 - 70 K/mW, we judge that the temperature
rise should be less than 4\ K. Moreover, any suggested decrease of
the superconducting gap caused by heating should be gradual and the
corresponding $I-V$ feature should not be sharp.  However, Joule
heating increases with current and gradually becomes visible at higher bias
currents, as will be shown below.

\section{High-current measurements}

As it is seen in Fig.~\ref{IVC}, the simulated $I-V$ curve
has no back-bending in contrast to what is seen in the experiment.
In fact, the back-bending in the experiment is so large that the
corresponding voltage eventually becomes zero at high current. We
think that both the back-bending and the current-induced zero-voltage
state can be qualitatively explained by Joule heating in the region
where the current is supplied to the junction.

\par

Indeed, the heat dissipation and, correspondingly, the temperature
of the stack progressively increase with the bias current. The
temperature dependence of the bridge resistance $R_{ab} = R_{ab}(T)$
and its critical current $I_c = I_c(T)$ should then be taken into
account. The former increases, while the latter decreases with
temperature. Qualitatively, this means that increasing the bias
current will result in a reduction of its fraction that is branching
off towards the potential electrode. In other words, a smaller
current is flowing through J2 for higher current biases. Even if
taking into account the somewhat compensating effect of the
nonlinear $I-V$ characteristic in the $c$-axis direction, this can
still result in the smaller voltage drop across $R_{J2}$.
Eventually, when this current becomes smaller than the re-trapping
current for J2, it will switch into the zero-voltage state.

\par

Numerical simulations of the heating effects were done by including
a dependence of  $R_{ab}$ and $I_{c \parallel ab}$ on the current
through J1, $I_{J1}$, since the small stack under the current lead
is believed to be the main source of heating. The increase of
$R_{ab}(I_{J1})$ and decrease of $I_{c \parallel ab}(I_{J1})$ with
temperature could be tested with different dependencies. The curve
b) in Fig.~\ref{IVC} was obtained by assuming linear dependence.
Taking account of the real energy dissipation, i.e. assuming that
$R_{ab}$ and $I_{c \parallel ab}$ are functions of the product
$I_{J1}V_{J1}$ or $I_{J1}^2$ and not simply $I_{J1}$ did not yield
much better fits.

\par

It is worth noting that in order to get back-bending of the $I-V$
curve using this simple equivalent circuit, it is essential to
include the heating effects.  No form of non-linearity in
$V_{ab}(I_{ab})$ could help model the back-bending unless we
included the dependence $V_{ab}$ on $I_{J1}$, $V_{ab} =
V_{ab}(I_{ab}, I_{J1})$, as we shortly described above. Curves
resembling curve c) of Fig.~\ref{IVC} could only be obtained
otherwise in most cases.

\par

One should not expect quantitatively correct results from these
simulations due to the much simplified equivalent circuit used and
also due to a somewhat uncertain current distribution in the base
crystal below the SIJJ. The plane just below the mesa can switch to
the normal state at sufficiently high current as well. Then, the
voltage drop will also depend on the geometry of base current- and
potential contacts along the crystal surface. To take even this into
account looks unjustified in view of the simplifications which were
used.

\section{Multiple-junction stacks}

\begin{figure}
\includegraphics[width=7.5cm]{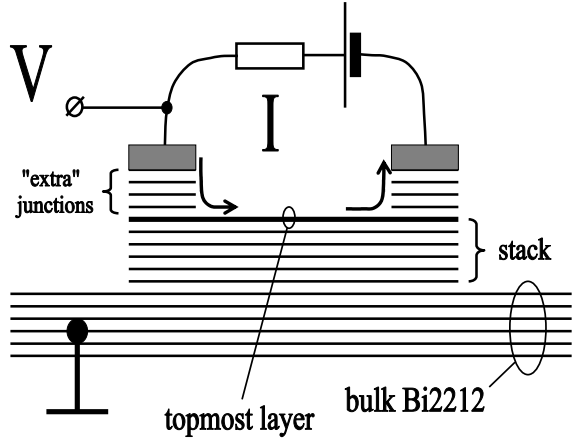} \caption{
A simple equivalent circuit of the multi-junction stack in
three-probe measurement when the bias current is fed into the
topmost layer of the stack through a few extra junctions under the
metallic contacts.  The voltage is measured relative to the bulk of
the Bi2212 single crystal using another mesa somewhere else on the
surface of the crystal.} \label{schematics}
\end{figure}

To see the $ab$-plane superconducting transition of a single plane
in higher stacks containing many junctions,  one can drive current
through the topmost electrode by applying the current between the
two top electrodes and measuring the voltage on either of the
contacts relative to the bulk of the single crystal, as is
schematically shown in Fig.~\ref{schematics}

\begin{figure}
\includegraphics[width=7.5cm]{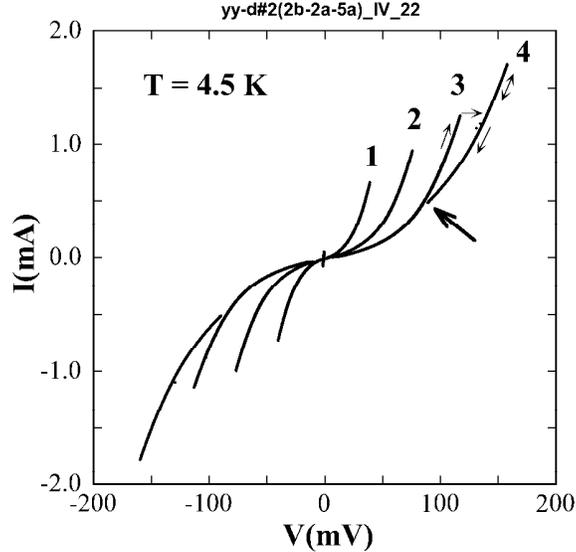} \caption{
A current-voltage characteristic of a mesa $\approx 30 \times 7\
\mathrm{\mu m}^2$ in area measured in accordance with schematics
shown in Fig.~\ref{schematics}. There are three ``extra" junctions
under each of the contacts seen as  three first branches
($1-3$)\cite{NJ}.  The ``pedestal" stack contains four junctions
($I-V$ not shown). The zero-voltage superconducting branch has a
much smaller critical current due to the topmost ``extra" junction
is in direct contact to the normal-metal thin film.  The thick arrow
marks the position of the break corresponding to the superconducting
transition of the surface layer when the current through it becomes
less than the critical value. Thin arrows indicate current
directions for tracing out the hysteresis in $I-V$ characteristic.}
\label{break}
\end{figure}

In the arrangement without ``extra" junctions, no voltage is
expected to appear unless the in-plane current along the topmost
layer exceeds the critical value. If dealing with the U-shaped mesas
with a certain number of "extra" junctions, the voltage will be
equal to the voltage across these junctions (plus small voltage drop
across the contact resistance). A feature corresponding to the
current-driven transition to the normal state of the topmost
electrode is a break of the last quasiparticle branch. The break is
also accompanied by the appearance of an extra $I-V$ branch, see
Fig.~\ref{break}.

In this figure, an $I-V$ curve of a stack having two contacts on top
is shown, with current directed from one to another. The voltage is
measured relative to the bulk using another mesa somewhere else on
the crystal surface.

There are four quasiparticle branches seen, three having a common
point of origin at $V,I \rightarrow 0$ and the last one having some
offset both in voltage and current for the point of origin. The
latter is marked by the thick arrow.  We argue that this point
corresponds to the current-driven superconducting transition of the
topmost layer of the stack. Let us describe the whole picture in
detail.

The bias current first flows through the small stack (extra
junctions) under one contact, then along the topmost electrode and
finally, through the small stack on the other end of the mesa. There
are three extra IJJ's under each of the contacts on top of a larger
stack (pedestal) which contains four junctions. The latter can be
seen in the four-probe measurements when the current is directed
vertically down to the bulk and the voltage is measured at the
second contact of the mesa, exactly as it was done in the case of
the single-junction measurements, see above.

When the current exceeds the critical value for the plane, the
current re-distributes, with one part still flowing along the plane,
and the rest flowing right through the junction under the bottom of
the extra-junctions stack. At this moment however, no additional
voltage appears since that junction is still in the superconducting
tunneling state. The current which is branching off downwards must
become larger than the critical current for that ``hidden" junction.

When it happens, one more extra junction becomes visible, see branch
``4" in Fig.~\ref{break}. This means that in the case of ideally
equivalent (``extra") junctions in the small stack, the observed
critical current for branch 3 should be larger than the critical
current for the previous branches by the value of the in-plane
critical current that is branching off towards another electrode.

If the current is set to decrease after that, the hidden junction
will not leave the quasiparticle-tunneling state unless the current
through it becomes less than the re-trapping current. This means
that when the \textit{total} bias current becomes less than the sum of
the in-plane
critical current and the re-trapping current of that hidden junction,
the system switches back to the state corresponding to branch 3.

Following this scenario, the critical current of the topmost single
$\mathrm{Cu_2O_4}$ plane corresponds to the current at which
branches 3 and 4 cross, see the arrow in Fig.~\ref{break}, because
the re-trapping current for a typical IJJ is of the order of few tens of
$\mu$A, i. e. much smaller than the in-plane critical current. Given the
width of the stack $w=7\ \mu$m, we calculate the sheet critical
current density to be $\approx 0.7\ \mathrm{A/cm}$ corresponding to
the bulk in-plane current density of 4.7~MA/cm$^2$.

The superconducting transitions of the second and even third from
the top layer can be seen at high enough current as well. At high
current however, the Joule heating progressively increases and makes
estimations of the corresponding critical current densities somewhat
understated.

\section{Discussion}

We have argued that the sheet critical current density of a single
Cu-O plane corresponds to the critical current density which would
be obtained in a perfect bulk sample. However, the critical current
of an isolated superconducting plane in BSCCO should be limited by
large thermal fluctuations associated with the two-dimensional (2D)
character of the Cu-O planes. Topological defects in the form of 2D
vortex-antivortex pairs are likely to arise as a result of these
fluctuations, as first suggested by Berezinskii, Kosterlitz, and
Thouless (BKT). \cite{Berezinskii:JETP72,Kosterlitz:JPC73} Long
enough vortex dipoles can be broken by the transport current and the
resulting free 2D vortices will sweep across the superconductor
giving rise to dissipation in the system.  This scenario leads to a
non-linear behavior of the $I-V$ characteristics since the number of
free vortices obviously depends on the bias current.

\par

We can calculate the sheet critical current
density assuming that the in-plane critical current is
limited by the BKT-mechanism:\cite{Jensen:PRL91}
\begin{equation} \label{eq2}
i_c \left[\mathrm{ \frac {A}{cm}}\right] = 5.8 \times 10^8 \frac {1} {\lambda_{ab}^2
[\AA]} \sqrt {\frac{m_{ab}}{m_c}}
\end{equation}
Where $\lambda_{ab}$ is the London penetration length; $m_{ab}$ and
$m_c$ are the effective mass of the superfluid particle for motion
in $ab$ plane and $c$ direction separately.

\par

Taking $\lambda_{ab} \approx 1800$ \AA\ and $m_c / m_{ab} \approx 2
\times 10^5$, Eq.\ref{eq2} yields $I_c \approx 0.4$ A/cm which is
very close to the experimental values of 0.3-0.7 A/cm. This is also
consistent with early suggestions on BKT-type phase transition of
the layered HTS. \cite{Jensen:PRL91,Glazman:JETP90,Artemenko:PLA90}

\par

Still, the values that we obtain, are among the largest ever
observed for BSCCO, both in thin films
\cite{Yonemitsu:PhysicaC02,Moriya:SUST02}, single crystals
\cite{Li:PhysicaC96,Matsushita:SUST98,Latyshev:PhysicaC93,Mizutani:Physica03},
and tapes \cite{Flahaut:IEEEASC03,Marken:IEEEASC01}.  Remarkably,
our results are even larger than for whiskers
\cite{Latyshev:PhysicaC93,Mizutani:Physica03}. Whiskers are believed
to represent the highest quality of BSCCO single crystals, having
virtually no stacking faults and grain boundaries. Such extended
defects are anticipated to limit the bulk critical current.   To the
best of our knowledge however, the largest value of $J_{c\parallel
ab}$ reported for whiskers \cite{Latyshev:PhysicaC93} is
0.5~$\mathrm{MA/cm^2}$, i.e. a few times smaller than in our
experiments and than what can be deduced using Eq.~2. We believe
that the failing to inject current uniformly through the whiskers'
cross section may result in large underestimation of the current
density. In most cases, the contacts to the whiskers were applied
from the flat in-plane sides causing a highly-nonuniform current
distribution\cite{Bush:PRL92}.

\par

The breaks in $I-V$ curves similar to the one presented in
Fig.~\ref{break} have also been seen in several other studies of
multi-junction stacks of intrinsic Josephson
junctions\cite{Doh:PRB00,Irie:PRB00,Saito:APL04,Krasnov:PRB02}.
However, the authors neither commented on these features nor
explained them.

\par

The position of this feature in current depends, of course, on sizes
of the stack, the perimeter around it through which the current
spreads out over the surface towards another contact on the single
crystal, and on superconducting properties of the particular single
crystal used. Some geometries are especially favorable for seeing
the break in $I-V$'s, like the case of having one stack placed on
top of another one. This is the geometry of our experiment described
above while similar arrangements can be found in other published
experiments as well\cite{Doh:PRB00,Krasnov:PRB02}. In our geometry,
the current spreads out over the surface of the base (pedestal)
stack through one-fourth of the square perimeter of the small stack
sitting on top of the former stack, see Figs.~\ref{SIJJ} and
\ref{schematics}. It is clear that in the case of a stand-alone
mesa, the current of the break should be four times larger.

The current-induced collapse of superconductivity of the surface
layers can have a profound effect on several relatively high-bias
tunneling-spectroscopy measurements on Bi-family of high-temperature
superconductors. These include  the break-junction technique and the
so called intrinsic-tunneling spectroscopic measurements, in
particular.

Indeed, the large superconducting energy gap of high-temperature
superconductors requires quite high current densities to reach the
corresponding gap feature in such experiments. If the current
exceeds the critical value for the topmost plane of the base  crystal,
an extra
contribution from the "hidden" junctions or/and the in-plane
normal-state resistance will add to the measured voltage. How large
the false contribution is depends on the particular geometry of
experiment but qualitatively, it is anticipated to result in
overestimation of the superconducting energy gap.

\section{Conclusions}

The sheet critical current density of a \textit{single} Cu-O plane was
measured by forcing the current to flow along the topmost electrode
of a single intrinsic Josephson junction. The critical current was
marked by a steep upturn (break) in the quasiparticle branch of the junction(s)
when the superconductivity of the topmost Cu-O layer is destroyed by
current. Further increase of the bias current caused a back-bending
of the $I-V$ curve followed by a reentrance of the zero-voltage
state. This is explained by progressive increase of Joule-heating at
high bias.

The values of the sheet critical current, 0.3-0.7 A/cm are among
largest values ever reported for BSCCO and are close to
theoretically estimated ones assuming a critical-current-limiting
BKT nature of 2D superconductivity in cooper oxides.

\begin{acknowledgments}

We thank M. Torstensson and D. Lindberg for technical assistance,
T. Claeson and V. Krasnov for fruitful discussions. This work is
financed by The Swedish Foundation for Strategic Research (SSF)
through the OXIDE program.

\end{acknowledgments}


\begin{thebibliography}{99}
\bibitem{NJ} The $I-V$ curve of IJJs (Fig.~\ref{SIJJ}c) usually consists of several branches, each
corresponding to one, two, three, and so on, intrinsic Josephson
junctions switching to the quasiparticle-tunneling state with
increasing bias current. Therefore the height $h$ can be obtained by
multiplying the height of individual IJJ by the number of
quasiparticle branches of the $I-V$ characteristic (voltage measured
between $V_N^+$ and $V^-$), i.e. $h \sim N \times 1.5$ nm. In the
same way, the depth $d$ can be obtained by multiplying the height of
individual IJJ by the number of quasiparticle branches of the $I-V$
characteristic (voltage measured between $V_N^+$ and $V_1^-$).
\bibitem{contacts} In fact, other stacks with low contact resistances and much
larger areas are used for this purpose. Making contacts to the
etched areas can be problematic.  The low contact resistances are
only possible to reproducibly acheive when the gold thin film is
deposited as soon as possible after cleaving the single crystal.
\end{thebibliography}
\end{document}